\documentclass[runningheads]{svmult}

\usepackage{makeidx}   
\usepackage{graphicx}  
\usepackage{subeqnar}  
\usepackage{multicol}  
\usepackage{physprbb}  
\makeindex             

\def\sqr#1#2{{\vcenter{\hrule height.#2pt\hbox{\vrule width.#2pt
height#1pt \kern#1pt \vrule width.#2pt}\hrule height.#2pt}}}
\def\square{\mathchoice\sqr64\sqr64\sqr{4.2}3\sqr{3.0}3}

\begin{document}
\title*{How does the electromagnetic field couple to gravity, in
    particular to metric, nonmetricity, torsion, and
    curvature?}
\toctitle{How does the electromagnetic field couple to gravity, in
    particular to metric, nonmetricity, torsion, and
    curvature?}
\titlerunning{How does the electromagnetic field couple to gravity?}
\author{Friedrich W. Hehl\inst{1}\inst{2}
\and Yuri N. Obukhov\inst{3}}
\authorrunning{Friedrich W. Hehl and Yuri N. Obukhov}
\institute{School of Natural Sciences\\
Institute for Advanced Study\\
Princeton, NJ 08540, USA
\and Institute for Theoretical Physics\\
University of Cologne\\
50923 K\"oln, Germany
\and Department of Theoretical Physics\\
Moscow State University\\
117234 Moscow, Russia}
\maketitle              
\begin{abstract}
  The coupling of the electromagnetic field to gravity is an age-old
  problem. Presently, there is a resurgence of interest in it, mainly
  for two reasons: (i) Experimental investigations are under way with
  ever increasing precision, be it in the laboratory or by observing
  outer space. (ii) One desires to test out alternatives to Einstein's
  gravitational theory, in particular those of a gauge-theoretical
  nature, like Einstein-Cartan theory or metric-affine gravity.--- A
  clean discussion requires a reflection on the foundations of
  electrodynamics. If one bases electrodynamics on the conservation
  laws of electric charge and magnetic flux, one finds Maxwell's
  equations expressed in terms of the excitation $H=({\cal D},{\cal
    H})$ and the field strength $F=(E,B)$ without any intervention of
  the metric or the linear connection of spacetime. In other words,
  there is still no coupling to gravity. Only the constitutive law
  $H={\rm functional}(F)$ mediates such a coupling. We discuss the
  different ways of how metric, nonmetricity, torsion, and curvature
  can come into play here. Along the way, we touch on non-local laws
  (Mashhoon), non-linear ones (Born-Infeld, Heisenberg-Euler,
  Pleba\'nski), linear ones, including the Abelian axion (Ni), and
  find a method for {\em deriving} the metric from linear
  electrodynamics (Toupin, Sch\"onberg). Finally, we discuss possible
  non-minimal coupling schemes. 
\end{abstract}

\section{Introduction}

General relativity was proposed in 1915. One of its predictions was
the bending of light rays of stars in the gravitational field of the
Sun. This effect was verified observationally soon afterwards by
Dyson et al.\ in 1920 and put, as a result, Einstein's theory in
the forefront of gravitational research.

Within the framework of general relativity, a light ray can be
extracted from classical electrodynamics in its {\em geometrical
  optics} limit, i.e., for wavelengths much smaller than the local
curvature radius of space. Accordingly, the bending of light can be
understood as a result of a nontrivial refractive index of spacetime,
see Skrotskii et al.\ \cite{Skro1,Skro2}, due to the coupling of the
electromagnetic field $F$ to the gravitational field $g$. Classically,
we have in nature just these two fundamental fields $F$ and $g$, the
weak and the strong fields being confined to microphysical dimensions
of $10^{-19}\rm m$ or $10^{-15}\rm m$, respectively. Therefore, the
coupling of $F$ and $g$ is of foremost importance in classical
physics.

The conventional way that coupling is achieved is to display the
Maxwell-Lorentz equations of vacuum electrodynamics in the (flat)
Minkowski world of special relativity theory in Cartesian
coordinates. For this purpose, usually the formalism of tensor
analysis (Ricci calculus) is used, see \cite{Schouten}:
\begin{equation}\label{Maxwell-L}
F^{ij}{}_{,j}=I^i\,,\qquad\quad F_{ij,k}+F_{jk,i}+F_{ki,j}=0\,.
\end{equation}
Here $F_{ij}=-F_{ji}=(F_{01},F_{02},F_{03},F_{23},F_{31},F_{12}
) =(\vec{E},\vec{B})$ is the electromagnetic field
strength, $I^i$ the electric 4-vector current, and
\begin{equation}\label{Fcontra}
F^{ij}:=g^{ik}g^{jl}\,F_{kl}\,,
\end{equation}
with $g^{ij}$ as the contravariant components of the metric. The
commas in (\ref{Maxwell-L}) denote partial differentiation
with respect to the local spacetime coordinates $x^i$.

If we switch on {\em gravity}, the flat Minkowski world becomes
curved, the spacetime geometry now being a Riemannian geometry with a
variable metric $g_{ij}(x)$ of Minkowskian signature $(+--\,-)$. The
coupling of the Maxwell-Lorentz set (\ref{Maxwell-L}) to gravity is
now brought about by the {\em comma goes to semicolon rule}
,$\;\rightarrow\;$; (see \cite{MTW}), where the semicolon represents
the covariant derivative $\nabla_i\equiv{}_{;i}$ with respect to the
Riemannian connection (``Levi-Civita connection''):
\begin{equation}\label{Maxwell-L;}
F^{ij}{}_{;j}=I^i\,,\qquad\quad F_{ij;k}+F_{jk;i}+F_{ki;j}=0\,.
\end{equation}
This translation rule from special to general relativity is also
alluded to as {\em minimal coupling}\index{Minimal coupling principle}
with the additional understanding that the components of the metric 
in (\ref{Fcontra}) become spacetime dependent fields.

The metric field $g_{ij}(x)$, entering (\ref{Maxwell-L;}) via
(\ref{Fcontra}) and via the covariant derivatives, i.e., via the
semicolons, has to fulfill the Einstein field equation\index{Einstein 
equation},
\begin{equation}\label{Einstein}
{Ric}_{ij}-\frac{1}{2}\,g_{ij}\,{Ric}_k{}^k=
\kappa\left(\stackrel{_{\rm Max}}{T}_{ij}+ \stackrel{_{\rm
mat}}{T}_{ij}\right)\,,
\end{equation}
with 
\begin{equation}\label{Einstein'}
{Ric}_{ij}:=R_{kij}{}^k\,,\qquad\stackrel{_{\rm
Max}}{T}_i{}^j:=\sqrt{\frac {\varepsilon_0}{\mu_0}}
\left(-\,F_{ik}F^{jk} + \frac{1}{4}\,
\delta_i^j\,F_{kl}F^{kl}\right)\,.
\end{equation}
Here $R_{ijl}{}^k$ is the curvature and $\stackrel{_{\rm
mat}}{T}_{ij}$ the material energy-momentum tensor. The coupled
Einstein-Maxwell system describes correctly a wealth of experiments,
in particular the gravitational bending of light, the gravitational
redshift, the time delay of radar pulses in the gravitational field of
the Sun, and the gravitational lensing and microlensing of starlight
in the gravitational field of galaxies.

But in all these experiments, we study the propagation of light along
null-geodesics in a prescribed (and perhaps slowly varying)
gravitational field which is a solution of the Einstein vacuum
equation -- and not of the {\em electro}-vacuum equation. We could
call this the {\em non} self-consistent Einstein-Maxwell theory. In
the solar system, e.g., the Schwarzschild metric is taken as solution
of the Einstein vacuum equation and the motion of a ``photon'' is
described by the null geodesic equation on this background. A true
novel effect of the Einstein-Maxwell theory would be, e.g., the
generation of electromagnetic waves by gravitational waves. Because of
their smallness, no such effects were ever observed. Accordingly, the
interaction of a classical electromagnetic field $F_{ij}$ in the form
of a lightray with a prescribed gravitational field $g_{ij}(x)$ is
well described by means of Eqs.(\ref{Maxwell-L;}). Nevertheless,
further consequences of these equations need to be compared with
experiment as soon as more sensitive measuring methods are available.

\section{On the equivalence principle}\index{Equivalence principle}

According to Einstein's equivalence principle, see \cite{Meaning},
gravity can be {\em locally simulated} in a gravity-free region of
spacetime by going over from the Cartesian coordinates, anchored in an
inertial frame of reference (including an intertial clock) and used in
(\ref{Maxwell-L}), to arbitrary curvilinear coordinates yielding a
non-inertial frame in general, as in (\ref{Maxwell-L;}). In this
context, the metric $g_{ij}$, occurring in (\ref{Fcontra}) and in the
semicolons of (\ref{Maxwell-L;}), is understood as a flat metric in
curvilinear coordinates. Thus, the minimal coupling\index{Minimal coupling 
principle} can be interpreted, in a first step, just as a coordinate 
transformation from Cartesian to curvilinear coordinates. And, moreover, 
it {\em identifies} the metric as the gravitational potential.

On the other hand, let us assume that we are in a region {\em with}
gravity and (\ref{Fcontra}) and (\ref{Maxwell-L;}) are valid together
with the Einstein equation for the metric. Then, also according to
Einstein's equivalence principle, we must be able to pick suitable
coordinates such that locally the equations look like in special
relativity in Cartesian coordinates. In Riemannian geometry, the local
coordinates are called Riemannian normal (hence geodesic) coordinates
at one point $P$, if the Christoffel symbols
\begin{equation}\label{Christ}
\Gamma_{ij}{}^k:=\frac{1}{2}\,g^{kl}\,\left(
g_{il,j}+g_{jl,i}-g_{ij,l}\right)
\end{equation} vanish at $P$ and and the metric becomes Minkowskian:
\begin{equation}\label{geodesic}
\Gamma_{ij}{}^k|_{\rm P} \stackrel{*}{=}0\,,\qquad g_{ij}|_{\rm P}
\stackrel{*}{=}{\rm diag}\;(+1,-1,-1,-1)\,.
\end{equation}
Accordingly, the semicolon becomes a comma and the metric in
(\ref{Fcontra}), at one given point, looks flat.

Still, the curvature is non-vanishing, of course: $R_{ijk}{}^l|_{\rm
P}\neq 0$. The equations look flat since they contain only {\em first}
derivatives. If they contained second derivatives, then the semicolons
goes to comma rule and its reverse would {\em not} work since on that
level not only the Christoffels enter but potentially also the
curvature which, in contrast to the Christoffels, is a tensor and
cannot be nullified by means of a suitable choice of coordinates. For
that reason, the minimal coupling procedure, being in this context an
expression of the equivalence principle, must be applied only to first
order differential equations. The safest thing is then to apply it, as
a rule, only on the level of a {\em Lagrangian}, since there
ordinarily only first-order expressions are allowed for. Non-minimal
couplings\index{Non-minimal coupling} of the gravitational field to 
electromagnetism have also been investigated, see Prasanna \cite{Prasanna71}, 
Buchdahl \cite{Buchdahl}, Goenner \cite{Hubert1}, and M\"uller-Hoissen
\cite{MH88.0,MH88.1,MH88.2}, for example, or for light rays in
non-minimally coupled theories, see Drummond and Hathrell
\cite{Drummond}, but the price one has to pay is to introduce a new
constant of nature; and there is no evidence for such a constant in
nature -- unless one takes the Planck length itself. We will come back
to these questions in Sec.7.

Therefore we can conclude that the equivalence principle and minimal
coupling work well for the Maxwell-Lorentz equations (\ref{Maxwell-L})
and that they lead to experimentally established equations.

\begin{footnotesize}
\subsubsection*{Wave equation for the electromagnetic field strength}

We hasten to add that, within the framework of the minimally coupled
Maxwell-Lorentz equations, we find 2nd derivatives if we derive the
wave equation for the electromagnetic field strength $F$\index{Wave 
equation for electromagnetic field strength} --- and then also curvature 
terms are expected to emerge. This is exactly what happens, as already 
found by Gordon \cite{Gordon} and Eddington \cite{Eddington}.

In the framework of exterior calculus (see Frankel \cite{Ted}), let us
consider the electromagnetic field strength 2-form $F = {\frac 1
  2}\,F_{ij}\,dx^i\wedge dx^j$.  In Maxwell-Lorentz vacuum
electrodynamics, it satisfies $dF =0,\ \varepsilon_0 d\,\,^\star\! F =
{\frac 1 c}\,J$. We denote the codifferential\index{Codifferential}
by $\delta:={}^\star d\,^\star$. Then we find, with the wave operator 
(d'Alembertian)
\begin{equation}\label{d'Alembertian}
\square :=\delta\,d + d\,\delta\,,
\end{equation} and by using the Maxwell-Lorentz equations, the 
wave equation 
\begin{equation}
\square F=\frac{1}{\varepsilon_0c}\,d\,\,{}^\star\! J\,,
\end{equation}
see \cite{Mohanty}. The left hand side of this equation, in terms of
components, can be determined by substituting (\ref{d'Alembertian}):
\begin{equation}
\square\,F = {\frac 1 2}\left(\nabla^k\nabla_k\,
F_{ij} + 2\,{Ric}\,{}_{[i}{}^k\,F_{j]k} - R^{kl}{}_{ij}\,F_{kl}
\right)dx^i\wedge dx^j\,.\label{DeltaF}
\end{equation} Accordingly, minimal coupling can lead to 
curvature terms of a prescribed form.\footnote{Of course, we could
have non-minimal coupling as, e.g., in $\square
F+\gamma\,\left(e_\alpha\rfloor e_\beta\rfloor R^{\alpha\beta}\right)
\wedge F=\frac{1}{\varepsilon_0c}\,d\,\,^\star\! J\,,$ see Sec.7.}
\end{footnotesize}

\section{A caveat}

Soon after general relativity had been proposed, it became clear, see
Einstein \cite{EinsteinMax}, that one can introduce as {\em auxiliary
variables} the densities
\begin{equation}\label{densities}
{\cal F}^{ij}:=\sqrt{-g(x)}\,g^{ik}(x)\,g^{jl}(x)\,F_{kl}\,,\qquad\quad
{\cal I}^{\,i}:=\sqrt{-g(x)}\,I^{\,i}\,,
\end{equation}
with $g(x):=\det g_{ij}(x)$, in terms of which the Maxwell-Lorentz
equations (\ref{Maxwell-L}) can be rewritten in a metric-free way as
\begin{equation}\label{Maxwell-dens}
{\cal F}^{ij}{}_{,j}={\cal I}^{\,i}\,,\qquad\quad
F_{ij,k}+F_{jk,i}+F_{ki,j}=0\,.
\end{equation}
Similarly, the charge conservation law $I^i{}_{;i}=0$ can be put in
the form
\begin{equation}\label{cc}
{\cal I}^i{}_{,i}=0\,.
\end{equation}
The metric enters only via the densities defined in
(\ref{densities}). In fact, if we started from the set
(\ref{Maxwell-dens}) in special relativity right away, then no comma
goes to semicolon rule would have been necessary: These equations are
generally covariant, they are valid in arbitrary curvilinear
coordinates, be it in the framework of special or general relativity
theory.

In the calculus of {\em exterior differential forms} (Cartan
calculus), see Frankel \cite{Ted}, these equations can be
formulated very succinctly. We introduce the electric current as odd
3-form\index{Electric current 3-form},
\begin{equation}\label{current3}
J:=\rho-j\wedge dt=\frac{1}{3!}\,J_{ijk}\,dx^i\wedge dx^j\wedge dx^k\,,
\end{equation}
the electromagnetic excitation as odd 
2-form\index{Electromagnetic excitation 2-form}
\begin{equation}\label{excitation2} 
H={\cal D} - {\cal H}\wedge dt=\frac{1}{2!}\,H_{ij}\,dx^i\wedge
dx^j\,,
\end{equation}
and the electromagnetic field strength as even 
2-form\index{Electromagnetic field strength 2-form}
\begin{equation}\label{field2}
F= B+ E\wedge dt=\frac{1}{2!}\,F_{ij}\,dx^i\wedge
dx^j\,.\end{equation}
Then (\ref{Maxwell-dens}) reads
\begin{equation}\label{Maxwell-excalc}
dH=J\,,\qquad\qquad dF=0\,, \end{equation}
with 
\begin{equation}\label{ccexcalc}
dJ=0\,. \end{equation} The set (\ref{Maxwell-excalc}) represents the
Maxwell equations\index{Maxwell equations}. They are independent of 
metric and connection. The constitutive relation for the 
vacuum\index{Constitutive law!vacuum} reads
\begin{equation}\label{vacuum} 
H=\sqrt{\frac{\varepsilon_0}{\mu_0}}\, {}^\star\! F\,,
\end{equation}
where the star $\star$ represents the metric-dependent and odd Hodge
duality operator. Eq.(\ref{vacuum}) corresponds to (\ref{densities})$_1$
and ${\cal I}^i$ can be related to the components of $J$,
\begin{equation}\label{transl}
{\cal F}^{ij}=\frac{1}{2!}\sqrt{\frac{\mu_0}{\varepsilon_0}}\,
\epsilon^{ijkl}H_{kl}\,, \qquad\qquad {\cal
I}^i=\frac{1}{3!}\,\epsilon^{ijkl}J_{jkl}\,,\end{equation} with
$\epsilon^{ijkl}=\pm 1,0$, the totally antisymmetric Levi-Civita
tensor density.

Now the equivalence principle looks empty: Since the Maxwell equations
(\ref{Maxwell-excalc}) are formulated in a coordinate and frame
independent way, they are valid in this form in arbitrary coordinate
systems and frames, be it in a flat or in a curved spacetime. Only the
constitutive relation (\ref{vacuum}) ``feels'', up to a conformal
factor, the presence of a flat or a non-flat metric, i.e., the
constitutive relation couples to the conformally invariant part of the
metric. The coupling of electromagnetism to gravity becomes almost
trivial. Is all this just a mathematical trick, which distracts from
the physical content of Maxwell's theory, or is it more?

One further observation hints also at the need for clarification. The
{\em Einstein-Cartan theory} of gravity is a {\em viable}
gravitational theory, see \cite{RMP,egg0,TrautmanMill}. It is the
simplest model of the metric-affine {\em gauge} theory of gravity, see
\cite{PRs,gron96}. In the Einstein-Cartan theory, spacetime is
described by means of a Riemann-Cartan geometry with torsion and
curvature.\footnote{A proper discussion of the equivalence principle
  in the context of Einstein-Cartan theory requires the introduction
  of local coframes, see \cite{gron96,hart,Iliev}. Being concerned
  here only with electromagnetism, it is sufficient to use natural,
  i.e., holonomic coframes.} If we couple (\ref{Maxwell-L;}) to
gravity, do we have to use the semicolons as covariant derivatives
with respect to the Riemann-Cartan connection or still with respect to
the Christoffels, see \cite{A-Pereira}?  In the context of
(\ref{Maxwell-excalc}), this question cannot even be posed, since the
exterior derivative $d$ is all what is needed. Are then the equations
(\ref{Maxwell-excalc}) misleading as a starting point for coupling to
gauge gravity? What could be the appropriate starting point?

Provided one formulates Maxwell's theory and its coupling to gravity
in terms of a Lagrangian with the electromagnetic potential $A$ as
variable, gauge covariance of the formalism results in
(\ref{Maxwell-excalc}) cum (\ref{vacuum}), as was pointed out by Benn,
Dereli, and Tucker \cite{BennT}. However, we would like to have some
more immediate insight into the structure of electromagnetism as
induced by experiment even without having a variational formulation at
our disposal.

\section{Electric charge and magnetic flux conservation}\label{First}

The metric is a quantity which allows to define {\em lengths} and {\em
angles} in spacetime. There are, however, laws in physics which don't
require the knowledge of a metric. Take the conservation law of
electric charge as an example. Mark a 3-dimensional simply connected
submanifold $\Omega_3$. We know from experiment that a possible electric
charge inside $\Omega_3$ is composed of charge ``quanta'', i.e., there
is an integer number of elementary charges in $\Omega_3$. Recent
advances in technology made it possible, see
\cite{Dehmelt,capacitance}, to trap and to count single electrons and
protons. Thus, as soon as we have such quanta available, we can rely
on {\em counting procedures}, see Post \cite{Post95}, the use of a
meter stick or a chronometer is superfluous under such circumstances.

Electric charge conservations is experimentally well-established and
is one of the pillars electromagnetism rests on. We formulate it,
following Kottler-Cartan-van Dantzig, see \cite{Post62,Truesdell} and
also \cite{HOR1}, most appropriately as an integral law. According
to (\ref{current3}), we assume the existence of the odd electric
current 3-form $J$\index{Electric current 3-form}. We take charge 
conservation as axiom 1\index{Electric charge conservation (Axiom 1)}, 
that is, $J$ integrated over a closed 3-dimensional hypersurface 
$\Omega_3$ has to vanish, if this hypersurface is the boundary of 
a connected 4-volume $\Omega_4$:
\begin{equation} \label{axiom1} 
\int\limits_{\partial\Omega_4}J=\int\limits_{\Omega_4}dJ=0\,,
\qquad{\rm or} \qquad dJ=0\,.
\end{equation} 
Here we applied the Stokes theorem.

If (\ref{axiom1}) is assumed to be valid $\int_{C_3}J=0$ for all 
three-cycles $\partial C_3 =0$, $C_3\neq\partial\Omega_4$, then, 
according to a theorem of de Rham, $J$ is exact, see \cite{Post95}. 
Thus the inhomogeneous Maxwell equation\index{Maxwell equations}
is a consequence,
\begin{equation}\label{JDH} 
J=d H\,, 
\end{equation}
with the odd 2-form $H$ of the electromagnetic excitation, see
(\ref{excitation2}).  The excitation is only determined up to an exact
form. Nevertheless, the electric excitation $\cal D$ can be {\it
measured} by means of Maxwellian double plates as charge per unit
area, the magnetic excitation $\cal H$ by means of a small test coil,
which compensates the $\cal H$-field to be measured, as current per
unit length. This is possible since in theses null experiments
vanishing field strength $F$ implies vanishing excitation $H$, see
\cite{HOR1}. In other words, the extensive quantities $\cal D$ and
$\cal H$ -- and thus the 4-dimensional excitation $H$ -- have an
operationally significance of their own, since they are related to
charge at rest or in motion, respectively.  Accordingly, the somewhat
formalistic introduction of the densities in (\ref{densities}) has now
been legitimized as a transition to operationally meaningful additive
quantities. Note that up to now only the differential structure of the
spacetime was needed, a metric has not been involved.

Let us choose a field of 4-frames $e_\alpha$ and consider the motion
of a point particle with respect to the reference frame thus
defined. As axiom 2 one can take an operational definition of the
electromagnetic field strength $F$ via the Lorentz 
force\index{Lorentz force density (Axiom 2)} density
\begin{equation}\label{axiom2}
f_\alpha=\left(e_\alpha\rfloor F\right)\wedge J\,.
\end{equation}
The interior product (contraction) is denoted by $\rfloor$. The force
density $f_\alpha$ is a notion from classical mechanics. It is an odd
covector-valued 4-form. Accordingly, Eq.(\ref{axiom2}) can be read as
a definition of the even 2-form $F$, see (\ref{field2}). Again, we
don't need a metric. And we know the recipe of how to proceed in the
same manner.

That $\int_{\Omega_2}F$ can be interpreted as magnetic flux is obvious
if we choose $\Omega_2$ as a `spacelike' surface (strictly, at this
point we don't know what spacelike means; we will come back to this
later). In superconductors under suitable circumstances we can count
(in an Abrikosov flux line lattice) quantized magnetic flux lines. This 
suggests that magnetic flux\index{Magnetic flux conservation (Axiom 3)} 
is a conserved quantity (axiom 3):
\begin{equation}\label{axiom3} 
\int\limits_{\partial\Omega_3}F=\int\limits_{\Omega_3}dF=0\qquad{\rm
or}\qquad dF=0\,.
\end{equation}

In this way, by means of the axioms (\ref{axiom1}), (\ref{axiom2}),
and (\ref{axiom3}), we recovered the fundamental structure of
Maxwell's theory: $dH=J,\> dF=0\,$\index{Maxwell equations}. 
This is what had been called 
{\em metric-free electrodynamics}.\footnote{Stachel \cite{Stachel} 
calls it {\em generalized} electrodynamics. We don't follow this 
suggestion, since Maxwell's equations were originally given in terms 
of $({\cal D},{\cal H})$ and $(E,\mu{\cal H})$ in a form `isomorphic' 
to the $(1+3)$-decomposition of $dH=J$ and $dF=0$, see \cite{Maxwell}. 
Therefore, the ``generalized'' Maxwell equations, $dH=J$ and $dF=0$, 
correspond in actual fact, just to {\em Maxwell's equations} (modulo 
the substitution $B\rightarrow\mu{\cal H}$). 
And this is how we will name them.}
What is missing so far is the relation between the excitation $H$ and
the field strength $F$, and it is exactly there where the metric,
i.e., the gravitational potential comes in.

\section[No interaction of ``substrata'' 
with gravity]{No interaction of charge and flux \\``substrata'' with
gravity}

We now understand that the inhomogeneous Maxwell equation $dH=J$, as
an expression of electric charge conservation, cannot be influenced by
gravity, i.e., by the metric tensor $g$, or, in the case of
metric-affine gravity or its specific subcases, such as
Einstein-Cartan theory, by the connection $\Gamma$ of spacetime. The
electric charge substratum of spacetime has rules of its own.
Spacetime can be deformed by the presence of metric and connection,
but the charge substratum and the net electric charge to be attributed
to a prescribed 3-dimensional (3D) volume won't change. Thus the
additivity 3D volume-wise of the charge lays at the foundation of the
Maxwellian framework. And it translates into the 2D additivity of the
integrated excitation $\int_{\Omega_2}H$ -- this being the reason why
one uses this integral for the operational interpretation of $H$.

Similar arguments can be advanced for the homogeneous Maxwell equation
$dF=0$. However, first of all it should be stressed that the
axiomatics we are using strongly suggests the non-existence of
magnetic charges. If there were magnetic charges, then we would have
no reason to believe in electric charge conservation either; compare
for this argument axiom 1, Eq.(\ref{axiom1}), with axiom 3,
Eq.(\ref{axiom3}). Conventionally, the inhomogeneous equation $dH=J$
is seen in analogy to the homogeneous one $dF=0$. But not so in the
framework of our axiomatics which has a firm empirical basis. We put
$dJ=0$ in analogy to $dF=0$. The whole historical development of
electromagnetism, starting with {\O}rsted and Amp\`ere, points to the
elimination of the phenomenologically introduced magnetic
charges. Most recent experiments, see \cite{Abbott,He}, exclude
magnetic charges with very good precision. Thus theoretical as well as
experimental evidence speak against the existence of magnetic charges.

Having said this, we hasten to add that, nevertheless, there is some
kind of magnetic substratum in spacetime, namely the magnetic flux
$\int_{\Omega_2}F$\index{Magnetic flux}. It is a substratum of its 
own right. The fluxoids, the quantized magnetic flux lines in 
superconductors, see \cite{Tinkham}, do convey a clear message. 
Besides electric charge\footnote{SI-unit Coulomb, elementary charge 
$e=1.60217733\times 10^{-19}\;{\rm C}$.}, magnetic flux\footnote{SI-unit 
Weber, elementary fluxoid $h/(2e)=2.06783461 \times 10^{-15}\;{\rm Wb}$.}
(and {\em not} magnetic charge) has an independent standing in
electromagnetism, too. Thus rightfully, it is governed by an own
axiom, namely axiom 3.

Axiom 3 is again a conservation theorem. In contrast to axiom 1, which
has a fermionic smell, axiom 3 is more of a bosonic nature. Moreover,
magnetic flux adds up 2D area-wise. For this reason, magnetic flux is
represented by a 2-form and not, like the charge, by a
3-form. Accordingly, there are essential differences between these two
conservation laws which express the peculiarities of the
electromagnetic phenomena. Electric and magnetic effects enter the
Maxwellian framework in an asymmetric way, in spite of all that talk
about a duality between electricity and magnetism. But there is also a
similarity in that both axioms are formulated as integral conservation
laws. The possibility to count the fluxoids assures us that axiom 3
has to be again a law free of metric and connection.

Incidentally, there is a nice visualization of the fundamental
quantities entering electrodynamics. If one describes the quantum Hall
effect for low lying Landau levels, then the concept of a {\em
  composite fermion} is very helpful: it consists of one electron and
an even number of fluxoids is attached to it, see Jain
\cite{Jain1,Jain2} and \cite{Johnson}.  Isn't that a very clear
indication of what the fundamental quantities are in electrodynamics?
Namely, electric charge (see axiom 1) and magnetic flux (see axiom 3),
see also Nambu \cite{Nambu} in this general context.

Our conclusion is then that, as long as we opt for electric charge and
magnetic flux conservation, the Maxwell equations in gravity-free
regions, i.e., in the Minkowski spacetime of special relativity, read
$dH=J\,\,{\rm and}\,\, dF=0$; they {\em remain the same} irrespective
of the switching on of gravity, be it in Einstein's theory, in
metric-affine gravity (see \cite{Punt}), or in any other geometrical
theory of gravity.

\section[Constitutive law and gravity]{Constitutive law of 
electrodynamics and its relation to gravity}

After having discussed extensively that gravity does not influence the
Max\-well equations, we eventually turn to the constitutive law via
which gravity does influence electrodynamics. It is true, the charge
substratum and the flux substratum themselves do not couple to
gravity, as we have shown in the last section. However, the
interrelationship between both substrata is affected by
gravity. Metaphorically speaking, the ``flow'' of {\em each} of the
substrata is ruled by a particular gravity-independent conservation
theorem, but the flows of electric charge and magnetic flux are
coupled via a gravity-dependent constitutive law since, in the end,
magnetism has to be expressed in terms of electricity.

Let us choose arbitrary local spacetime coordinates $x^i$. Then we have,
\begin{equation}
H = {\frac 1 2}\,H_{ij}\,dx^i\wedge dx^j,\qquad 
F = {\frac 1 2}\,F_{ij}\,dx^i\wedge dx^j.\label{localHF}
\end{equation}

We will turn first to the electrodynamics of material media in order
to develop some intuition on the concept involved, but eventually, it
will be the {\em vacuum}, be it in inertial or non-inertial frames,
which will occupy the center of our interest.

\subsection{Non-local}\index{Constitutive law!non-local}

Moving {\it macroscopic} matter defines a $(1+3)$-splitting of
spacetime specified by a well defined average 4-vector velocity field
$u$ which describes the congruence of worldlines of the flow of the
medium.  Such a vector field can be defined operationally from the
motion of matter as follows.

Let a 3--dimensional arithmetic space $R^3$ be equipped with the
coordinates $\xi^a, a=1,2,3$. We will use these coordinates (known as
{\it Lagrange} coordinates in continuum mechanics) as labels which
enumerate elements of a material medium. A smooth mapping
$x_{(0)}:\,R^3\rightarrow X_4$ into the spacetime defines a
3--dimensional space domain (hypersurface) $V$ which represents the
initial distribution of matter. In local spacetime coordinates, this
mapping (or labeling) is given by the four functions
$x^i_{(0)}(\xi^a)$. It should be preserved at any time, i.e.\ along
any worldline of a particular element its labels $\xi^a$ are constant.

Given the initial configuration $V$ of matter, we parameterize dynamics
of the medium by the ``time'' coordinate $\tau$ which is defined as
the proper time measured along an element's worldline from the
original hypersurface $V$. The resulting local coordinates
$(\tau,\xi^a)$ are usually called the normalized comoving
coordinates. Thus finally, the motion of matter is described by the
functions $x^i(\tau,\xi^a)$. Subsequently, we define the 4--velocity
vector field by
\begin{equation}
u:=\partial_\tau = \left({\frac {dx^i} {d\tau}}\right)_{\xi^a={\rm
const}}\, \partial_i\,.\label{udef}
\end{equation}
Evidently, a family of observers comoving with the matter is
characterized by the same timelike congruence $x^i(\tau,\xi^a)$. They
are making physical (in particular, electrodynamical) measurements in
their local reference frames which drift with the material motion.

One says that a medium, moving in general, has {\it dispersion}
properties when the electromagnetic fields produce non-instantaneous
polarization and magnetization effects. The most general {\em linear}
constitutive law is then given, in the comoving system, by means of
the integral
\begin{equation}\label{non-local}
H_{ij}(\tau,\xi) = {\frac 1 2}\int d\tau' K_{ij}{}^{kl}(\tau,\tau')\,
F_{kl}(\tau',\xi)\,.
\end{equation}
The coefficients of the kernel $K_{ij}{}^{kl}(\tau,\tau')$ are called
the response functions. We expect the metric to be involved in their
set-up. Their form is defined by the internal properties of matter and
by the motion of a medium.

Mashhoon \cite{Bahram} has proposed a physically very interesting
example of such a non-local electrodynamics in which non-locality
comes as a direct consequence of the {\em non-inertial} dynamics of
observers. In this case, instead of (\ref{localHF}), one should use
the field expansions
\begin{equation}
H = {\frac 1 2}\,H_{\alpha\beta}\,\vartheta^\alpha\wedge\vartheta^\beta,
\qquad F = {\frac 1 2}\,F_{\alpha\beta}\,\vartheta^\alpha\wedge 
\vartheta^\beta\label{non-localHF}
\end{equation}
with respect to the coframe of a non-inertial observer $\vartheta^\alpha 
= e_i{}^\alpha\,dx^i$. The constitutive law is then replaced by 
\begin{equation}\label{non-local1}
H_{\alpha\beta}(\tau,\xi) = {\frac 1 2}\int d\tau' 
K_{\alpha\beta}{}^{\gamma\delta}(\tau,\tau')\,
F_{\gamma\delta}(\tau',\xi)\,,
\end{equation}
and the response kernel in (\ref{non-local1}) is now defined by the
acceleration and rotation of the observer's reference system. It is a
constitutive law for the vacuum as viewed from a non-inertial frame of
reference. 

Mashhoon imposes an addititional {\it assumption} that the kernel is
of {\it convolution} type, i.e., $K_{\alpha\beta}{}^{\gamma\delta}
(\tau,\tau') = K_{\alpha\beta}{}^{\gamma\delta}(\tau - \tau')$. Then
the kernel can be uniquely determined by means of the Volterra
technique, and often it is possible to use the Laplace transformation
in order to simplify the computations.  Unfortunately, although
Mashhoon's kernel is always calculable in principle, in actual
practice one normally cannot obtain $K$ explicitly in terms of the
observer's acceleration and rotation.

Preserving the main ideas of Mashhoon's approach, one can abandon the
convolution condition. Then the general form of the kernel can be
worked out explicitly ($u$ is the observer's 4-velocity):
\begin{equation}
K_{\alpha\beta}{}^{\gamma\delta}(\tau,\tau') = {\frac 1
2}\,\epsilon_{\alpha\beta}{}^{\lambda[\delta}\!
\left(\delta^{\gamma]}_\lambda\,\delta(\tau -\tau') -
u\rfloor\Gamma_\lambda{}^{\gamma]}(\tau')\right).\label{NewAnsatz}
\end{equation}
The influence of non-inertiality is manifest in the presence of 
the connection 1-form. 
The kernel (\ref{NewAnsatz}) coincides with the original Mashhoon
kernel in the case of constant acceleration and rotation, but in
general the two kernels are different \cite{Muench}. Perhaps, only the
direct observations would establish the true form of the non-local
constitutive law. However, such a non-local effect has not been
confirmed experimentally as yet.

\subsection{Non-linear}\label{nonlin}\index{Constitutive law!non-linear}

But the constitutive law can also be non-linear (or non-local and
non-linear at the same time). In the local and non-linear {\em
Born-Infeld} electrodynamics \cite{BI}, with the dimensionfull
parameter $f_{\rm e}$ as maximal attainable electric field strength,
we have
\begin{equation}\label{non-linear1}
H=-\,\frac{\partial V_{\rm BI}}{\partial F}\sim
\frac{\partial\sqrt{-\,\det|g_{kl}+\frac{1}{f_{\rm e}}
F_{kl}|}}{\partial F}\,.
\end{equation}
The metric as symmetric second rank tensor enters here in a very natural 
way. It adds up with the antisymmetric electromagnetic field to an 
asymmetric tensor -- much in the way Einstein had hoped to find for his 
unified field theories of gravity and electromagnetism. 
By differentiation, we find\index{Born-Infeld electrodynamics}
\begin{equation}\label{BI}
  H = \sqrt{\frac {\varepsilon_0}{\mu_0}}\,
    \frac{{}^\star F + \frac{1}{2f_{\rm e}^2}\, {}^\star(F\wedge F)\, F}
    {\sqrt{1 - \frac{1}{f_{\rm e}^2}\,{}^\star(F\wedge{}^\star F) -
    \frac{1}{4f_{\rm e}^4}\,[{}^\star(F\wedge F)]^2}}\,,
\end{equation}
now the metric being absorbed in the (odd) {\em Hodge star} operator,
see \cite{Ted}. For $f_{\rm e}\rightarrow \infty$, we recover the
conventional local and linear Maxwell-Lorentz theory for vacuum with
$H=\sqrt{\frac {\varepsilon_0}{\mu_0}}\,{}^\star F$. The Born-Infeld 
electrodynamics is presently used as a toy model in string theories, 
see \cite{GiRa}. The problem with Born-Infeld electrodynamics is that, 
in contrast to Maxwell's theory, it defies quantization. It is an 
interesting model, but nothing like an established theory.

A similar example is the non-linear {\em Heisenberg-Euler}
electrodynamics \cite{HE}. Quantum electrodynamical vacuum
fluctuations yield corrections to Max\-well's theory that can be
accounted for by an effective constitutive law constructed by
Heisenberg and Euler. To the first order in the fine structure 
constant $\alpha_{\rm f}= {\frac {e^2} {4\pi\varepsilon_0\hbar c}}$, 
it is given by (see also \cite{Itzykson,heyl})
\begin{equation}\label{HE}
  H = \sqrt{\frac {\varepsilon_0}{\mu_0}}\left\{\left[
      1 + \frac{8\,\alpha_{\rm f}}{45\,B_k^2}\,
      {}^\star \bigl(F\wedge{}^\star  F \bigr)\right]\, {}^\star  F +
      \frac{14\,\alpha_{\rm f}}{45\,B_k^2} \,{}^\star \bigl(F\wedge
      F\bigr)\,F\right\},
\end{equation}
where $B_k={\frac {m^2c^2}{e\hbar}}\approx 4.4\times 10^9\,{\rm T}$, with 
the mass of the electron $m$\index{Heisenberg-Euler electrodynamics}. 
The metric is again hidden in the Hodge star and the Maxwell-Lorentz limit 
results analogously for $m\rightarrow\infty$. The Casimir force between 
two uncharged electrically conducting plates, also an effect of vacuum 
fluctuations, has been experimentally verified as have been non-linear 
effects in the ``superposition'' of strong laser beams. Accordingly, 
the non-linear constitutive law (\ref{HE}) is a valid post-classical 
approximation of vacuum electrodynamics and as such experimentally confirmed.

Note that these variants of classical electrodynamics respect charge
and flux conservation. This underlines the fact that our axiomatics
clearly points to that structure of electrodynamics, namely the
constitutive law, which can be changed without giving up the
essentials of electrodynamics.

Both, Eqs.(\ref{BI}) and (\ref{HE}) are special cases of Pleba\'nski's
more general non-linear electrodynamics \cite{Pleb}. Let the quadratic
invariants\index{Electromagnetic invariants} of the electromagnetic 
field strength be denoted by
\begin{equation}\label{Inv}
I_{1}:={\frac 1 2}{}^\star(F\wedge{}^\star F) = {\frac 1 2}(
{\vec{E}^2-\vec{B}^2})\quad {\rm and}\quad 
I_{2}:={\frac 1 2}{}^\star(F\wedge F) = \vec{E}\cdot\vec{B}\,,
\end{equation} 
where $I_{1}$ is an even and $I_{2}$ is an odd scalar (the Hodge
operator is odd). Then Pleba\'nski postulated a non-linear
electrodynamics with the constitutive law\footnote{Strictly,
  Pleba\'nski assumed a Lagrangian which yields (\ref{Maxwell-excalc})
  together with the {\em structural relations} $F=u(I_1,I_2)\,^\star H
  + v(I_1,I_2)\, H$. The latter law, apart from singular cases, is
  equivalent to (\ref{non-l}).}
\begin{equation}\label{non-l} H = U(I_{1},I_{2})\,{}^\star F+
V(I_{1},I_{2})\,F\,, \end{equation} where $U$ and $V$ are functions of
the two invariants. Note that in the Born-Infeld case $U$ and $V$
depend on both invariants whereas in the Heisenberg-Euler case we have
$U_{{\rm HE}}(I_1)$ and $V_{{\rm HE}}(I_2)$. Nevertheless, in both
cases $U$ is required as well as $V$. And in both cases, see
(\ref{BI}) and (\ref{HE}), $U$ is an even function and $V$ and odd one
such as to preserve parity invariance.

If one chose $V$ to be an even function, e.g., then parity violating
terms would emerge. Such terms were most recently discussed by
Majumdar, Mukhopadhyaya, and SenGupta \cite{Maj+Sen,Muk+Sen};$\>$ for
the experimental situation (there seem no signatures for parity
violations) compare Lue et al.\ \cite{Lue99}.

\subsubsection*{Singularity-free electro-gravitodynamics}

Recently, Ay\'on-Beato \& Garc{\'\i}a \cite{reg2}, for earlier work see
Shikin \cite{reg1}, have proposed a constitutive law
\begin{equation}\label{AG}
H = U(I_1)\,{}^\star F\,,
\end{equation}
which, as subcases, does neither encompass (\ref{BI}) nor (\ref{HE})
and thus makes it appear as rather academic. The explicit form of $U$
is {\it defined} by the requirement of obtaining completely
singularity-free solutions of the coupled system of the gravitational
field (Einstein) and the electromagnetic field (non-linear
Maxwell). Examples of suitable functions $U(I_1)$ are given in
\cite{reg1,reg2}.

In terms of the local time and space coordinates $(t, r, \theta,
\phi)$, the general spherically symmetric ansatz for the coframe can
be written as
\begin{equation}
\vartheta^{\hat 0} =\,f(r)\,d\,t\,,\quad\vartheta^{\hat 1} =\, 
{1\over f(r)}\,d\,r\,,\quad\vartheta^{\hat 2} =\, r\, d\,\theta\,,
\quad\vartheta^{\hat 3} =\, r\,\sin\theta \, d\,\phi\,,\label{frame1}
\end{equation}
whereas, for the electromagnetic field, we have 
\begin{equation}
F = \varphi(r)\,\vartheta^{\hat 0}\wedge\vartheta^{\hat 1}\,.\label{F1}
\end{equation}
The exact solution of the coupled system of gravitational and
electromagnetic field equations, i.e., of Einstein's equation
(\ref{Einstein}) and Maxwell's equations $dF =0,\,\, dH =0$, reads
\begin{equation}
\varphi = {\frac q {U(I_1)\,r^2}},\qquad
f^2 = 1 - {\frac {2m} r} + {\frac {Q(r)} {r^2}},\label{reg-sol}
\end{equation}
where $q, m$ are integration constants and (`Tolman's integral') 
\begin{equation}\label{QK}
Q(r) = \kappa\,r\,\int\limits^\infty_r dr'\,{\cal K}(r')\,{r'}^2,
\qquad {\cal K} = 2I_1\,U(I_1) - \int\limits^{I_1} dI_1'\,U(I_1').
\end{equation}
In the last function one should substitute the explicit form of the
quadratic invariant $I_1$ computed on the spherically symmetric
configuration (\ref{frame1}) with (\ref{F1}). 

It is shown in \cite{reg1,reg2,reg2a,reg3} that the constitutive
function $U(I_1)$ can be chosen in such a way that the functions in
(\ref{reg-sol}) describe a completely regular, i.e., singularity-free
configuration.

\subsection{Linear: Abelian axion, 
inter alia}\index{Axion}\index{Constitutive law!linear}

A very important case is that of a linear constitutive law between the
components of the two-forms $H$ and $F$.  It postulates the existence
of the $6\times 6=36$ constitutive functions $ \kappa_{ij}{}^{kl}(t,x)
=-\, \kappa_{ji}{}^{kl}=-\, \kappa_{ij}{}^{lk}$ such that
\begin{equation}
H_{ij} =  \frac{1}{2}\,\kappa_{ij}{}^{kl}\,F_{kl}.\label{chiHF}
\end{equation}
This kind of an ansatz we know from the physics of anisotropic
crystals. The factor $1/2$ is chosen in order to have a smooth
transition to the conventional
$\vec{D}=\varepsilon_0\,\varepsilon\,\vec{E}$ etc.\ relations, cf.\ 
\cite{Post62} p.127. 

The choice of the local coordinates is clearly unimportant. In a
different coordinate system the linear constitutive law preserves its
form due to the tensorial transformation properties of $
\kappa_{ij}{}^{kl}$.  Alternatively, instead of the local coordinates,
one may choose an anholonomic frame and may then decompose the
two-forms $H$ and $F$ with respect to it.

Since $H$ is an odd and $F$ an even form, the constitutive functions
$\kappa_{ij}{}^{kl}(t,x)$ are {\em odd}. Taking the Levi-Civita
symbol, we can split off the odd piece according to
\begin{equation}\label{lin1}
  \kappa_{ij}{}^{kl}=:\frac{1}{2}\,\epsilon_{ijmn}\,
 {\chi}^{mnkl}\qquad {\rm or}\qquad
 {\chi}^{ijkl}=\frac{1}{2}\,\epsilon^{ijmn}
\kappa_{mn}{}^{kl}\,.
\end{equation}
Because of the corresponding properties of the Levi-Civita symbol, 
the $\chi^{ijkl}$ are even scalar densities of weight $+1$.
For the Levi-Civita symbols with upper and lower indices, we have
$\epsilon^{ijkl}\,\epsilon_{mnpq}=\delta_{mnpq}^{\,\,i\,jk\,l}$.

With the linear constitutive law (as with more general laws), we can
set up a Lagrangian 4-form; here we call it $V_{\rm lin}$.  Because of
$H=-\,\partial V_{\rm lin}/\partial F$, the Lagrangian must be
quadratic in $F$.  Thus we find
\begin{eqnarray}\label{Vlin}
  V_{\rm lin}&=&-\,\frac{1}{2}\,H\wedge
  F=-\,\frac{1}{8}\,H_{ij}F_{pq}\,dx^i\wedge dx^j\wedge dx^p\wedge
  dx^q\nonumber\\
  &=& -\,\frac{1}{32}\,\left( \epsilon^{pqij}\epsilon_{ijmn}
    \chi^{mnkl}\right)F_{kl}F_{pq}\,dx^0\wedge dx^1\wedge
  dx^2\wedge dx^3\,.
\end{eqnarray}
The components of the field strength $F$ enter in a {\em symmetric}
way.  Therefore, without loss of generality, we can impose the
symmetry condition $\chi^{ijkl} = \chi^{klij}$
on the constitutive functions reducing them to 21 independent
functions at this stage.

The $\kappa_{ij}{}^{kl}$ carry the dimension $[\kappa]=[{\chi}]=
e^2/\hbar$. Therefore, still before introducing the metric, we can 
split off the totally antisymmetric part of $\chi^{ijkl}$ and define 
the dimensionless constitutive functions according to 
\begin{equation}\label{split}
\chi^{ijkl}=f\,\stackrel{\rm o}{\chi}{}^{\!ijkl}+\alpha\,\epsilon^{ijkl}\,,
\qquad{\rm with}\qquad \stackrel{\rm o}{\chi}{}^{\![ijkl]}=0\,.
\end{equation}
Here $[f] =[\alpha]=\hbar/e^2$, and $f =f(t,x)$ and
$\alpha=\alpha(t,x)$ represent one scalar and one pseudo-scalar
constitutive function, respectively.  Thus the {\em linearity ansatz}
eventually reads
\begin{equation}\label{linear}
H_{ij}= \frac{1}{4}\,
\epsilon_{ijmn}\,\chi^{mnkl}\,F_{kl}= \frac{f}{4}\epsilon_{ijmn}
\,\stackrel{\rm o}{\chi}{}^{\!mnkl}\,F_{kl}+\alpha\,F_{ij}\,,
\end{equation}
with 
\begin{equation}\label{chisymm}
\stackrel{\rm o}{\chi}{}^{\!mnkl}=-\stackrel{\rm o}{\chi}{}^{\!nmkl}
=-\stackrel{\rm o}{\chi}{}^{\!mnlk}=\,\stackrel{\rm o}{\chi}{}^{\!klmn}
\qquad{\rm and}\qquad \stackrel{\rm o}{\chi}{}^{\![mnkl]}=0\,,
\end{equation}
i.e., besides $\alpha$, we have 20 independent constitutive
functions. Thus $\stackrel{\rm o}{\chi}{}^{\!nmkl}$ has the same
algebraic symmetries and the same number of independent components as
a curvature tensor in a Riemannian spacetime.

Pseudo-scalars are also called axial scalars. So far, our axial scalar
$\alpha(x)$ is some kind of permittivity/permeability field. If one
adds a kinetic term of the $\alpha$-field to the electromagnetic
Lagrangian (\ref{Vlin}), then $\alpha(x)$ becomes propagating and one
can call it legitimately an Abelian\footnote{In contrast to the axions
related to {\em non}-Abelian gauge theories, see
\cite{Weinberg,Wilczek1,Moody} and the reviews in \cite{Kolb} and
\cite{Sikivie}.}  axion\index{Axion}. Ni \cite{Ni73} was the first 
to introduce such an axion field $\alpha$ in the context of the 
coupling of electromagnetism to gravity, see also deSabbata \& Sivaram
\cite{deSabbata1} and the references given there.

The {\em Abelian} axion has the following properties:
\begin{itemize}
\item Pseudoscalar field, i.e., spin $= 0$, parity $= -1$.

\item Couples to Maxwell's field in the Lagrangian according to
$\alpha\,F\wedge F=2\alpha\, E\wedge B\wedge dt$, see (\ref{linear})
and (\ref{Vlin}). Here $E$ is the 3-dimensional electric field 1-form
and $B$ the corresponding magnetic field 2-form. This term in the
total Lagrangian can be written as
\begin{equation}
\alpha\,F\wedge F = -\,d\alpha\wedge A\wedge F\,,
\end{equation}
dropping, as usual, the total derivative. This contributes to the
excitation $H = - \partial L/\partial F$ a term
\begin{equation}
\sim d\alpha\wedge A\,.\label{ax-mom}
\end{equation}

\item Since it arises on the same level as the metric, see
  Eq.(\ref{isotropy1}) below, it is a field of a similar universality
  as the gravitational field.
\end{itemize}

As yet, the Abelian axion has not been found experimentally, see the
discussion of Cooper \& Stedman \cite{Cooper} on corresponding ring
laser experiments.

\subsection{Isotropic}

The linearity ansatz (\ref{linear}) can be further constrained in
order to arrive eventually at an isotropic constitutive tensor. We
will proceed here somewhat unconventional in that we don't assume a
metric of spacetime beforehand but rather derive it in the following
way:

\subsubsection{Duality operator, electric and magnetic reciprocity} 

The constitutive tensor $\stackrel{\rm o}{\chi}{}^{\!klmn}$ of
(\ref{linear}) defines a new duality operator\index{Duality operator}
which acts on 2-forms on $X$. In components, an arbitrary 2-form 
$\Theta=\frac{1}{2} \Theta_{ij}\, dx^i\wedge dx^j$ is mapped into 
the 2-form ${}^\#\Theta$ by
\begin{equation}\label{dual} 
 {}^\#\Theta_{ij}:=\frac{1}{4}\epsilon_{ijkl}\,  
 {\stackrel{\rm o}{\chi}}{}^{klmn}\, \Theta_{mn}\, , 
\end{equation} see \cite{OH,HOR2}. 
No metric is involved in this process.  Now the linear material law
(\ref{linear}) can be written as
\begin{equation} 
H=\left(f\,{}^\# +\alpha \right)F. 
\end{equation} 
 
We postulate that the duality operator, applied twice, should, up to a
sign, lead back to the identity. Such a closure relation or the
``electric and magnetic reciprocity'' \cite{Toupin} reduces the number
of independent components of $\stackrel{\rm o}{\chi}$ to 9 (without
using a metric). One can demonstrate that this is a sufficient
condition for the {\em non}existence of {\em birefringence} in vacuum,
see \cite{Ni73,Ni77,Claus,Ni99,HLaem}. Then the fourth-order general
Fresnel equation degenerates to the second-order light cone equation.
Therefore, we impose\index{Closure relation}
\begin{equation}\label{cr} 
 {}^{\#\#}=-1 \, .  
\end{equation}  
The minus sign yields Minkowskian signature\footnote{One could define
a different duality operator by ${}^{\widehat\#}\Theta_{ij}=
\frac{f}{4}\epsilon_{ijkl}\,{\stackrel{\rm o}{\chi}}{}^{klmn}\,
\Theta_{mn}$ such that ${}^{\widehat{\#}\widehat{\#}}=-f^2$.},
whereas the condition $^{\#\#}= +1$ would lead to Euclidean or to the
mixed signature $(+,+,-,-)$.
 
Seemingly Toupin \cite{Toupin} and Sch\"onberg \cite{Schoen71} were
the first to deduce the conformally invariant part of a spacetime
metric from duality operators and relations like (\ref{dual}) and
(\ref{cr}).  This was later rediscovered by Jadczyk \cite{Jadczyk},
whereas Wang \cite{Wang} gave a revised presentation of Toupin's
results. A forerunner was Peres \cite{Peres}, see in this context also
the more recent papers by Piron and Moore \cite{Piron95}.
Brans \cite{Brans71} and subsequently numerous authors discussed such
structures in the framework of general relativity theory, see, e.g.,
\cite{Capo89,tHooft91,Harnett1,Tert} and the references given there.

It is convenient to adopt a more compact {\it bivector} notation by
defining the indices $I,J,\dots = (01, 02, 03, 23, 31, 12)$. Then
${\stackrel{\rm o}{\chi}}{}^{ijkl}$ becomes the $6\times 6$
matrix $\stackrel{\rm o}{\chi}{}^{IK}$ and (\ref{cr}) goes over into
\begin{equation}\label{cr2} 
 \stackrel{\rm o}{\chi}{}^{IJ}\,\epsilon_{JK}\stackrel{\rm
o}{\chi}{}^{KL}\, \epsilon_{LM}=-\delta^I_M \,.
\end{equation} 
In terms of $3\times 3$-constituents an arbitrary symmetric 
$\stackrel{\rm o}{\chi}{}^{IK}=\stackrel{\rm o}{\chi}{}^{KI}$ 
constitutive matrix reads
\begin{equation}  
 \stackrel{\rm o}{\chi}{}^{IJ}=\stackrel{\rm o}{\chi}{}^{JI}=  
 \left(\begin{array}{cr}A& C \\C^{{\rm T}} &B 
 \end{array}\right),\quad \epsilon^{IJ} =\epsilon^{JI}=  
 \left(\begin{array}{cr}0& {\mathbf 1} \\{\mathbf 1} 
&0\end{array}\right)\, , 
\end{equation}  
where $A=A^{\rm T}\,,B=B^{\rm T}$, and the superscript ${}^{\rm T}$
denotes transposition. The algebraic condition $\epsilon_{IJ} 
\stackrel{\rm o}{\chi}{}^{IJ}\equiv 0$ is provided by ${\rm tr}\,C =0$.

The general non-trivial solution of the closure relation (\ref{cr2})
can be written in the form
\begin{equation}\label{??} 
 \stackrel{\rm o}{\chi}{}^{IJ}=\left(\begin{array}{cc}pB^{-1} + qN& 
B^{-1}K \\  
 -KB^{-1} &B \end{array}\right)\, . 
\end{equation}
Here $B$ is a nondegenerate arbitrary {\em symmetric} $3\times 3$
matrix (6 independent components $B_{ab}$), $K$ an arbitrary {\em
antisymmetric} matrix (3 independent components $K_{ab}=:
\epsilon_{abc}\, k^c$), $N$ the symmetric matrix with components
$N^{ab}:=k^ak^b$, and $q:= - 1/{\det B}$, $p:=[{\rm tr}(NB)/\det B] -
1$. Thus, Eq.(\ref{??}) subsumes 9 independent components.

\subsubsection{Triplet of self-dual 2-forms}

The duality operator ${}^\#$ induces a decomposition of the
6-dimensional space of 2-forms into two 3-dimensional invariant
subspaces corresponding to the eigenvalues $\pm i$. Writing the 
2-form basis $\Theta^I=dx^i\wedge dx^j$ in terms of the two 
3-dimensional column vectors
\begin{equation} 
 \Theta^I = \left(\begin{array}{c} \beta^a \\ \gamma_b 
 \end{array}\right)\,,\quad a,b, \dots = 1,2,3, 
\end{equation}
one can construct the corresponding self-dual basis ${\stackrel 
{({\rm s})}\Theta}{}^I :=\frac{1}{2}(\Theta^I - i\,{}^\#\Theta^I)$.
In the 3-vector representation, 
\begin{equation} 
{\stackrel {({\rm s})} \Theta}{}^I = \left(\begin{array}{c}  
{\stackrel {({\rm s})} \beta}{}^a \\  
{\stackrel {({\rm s})} \gamma}_b \end{array}\right)\,,
\end{equation}
one of the 3-dimensional invariant subspaces can be spanned either by
the upper or by the lower components which are related to each other
by a linear transformation. For example, ${\stackrel {({\rm s})}
  \beta}$ can be expressed in terms of ${\stackrel {({\rm s})}
  \gamma}$ according to ${\stackrel{({\rm s})}\beta}=(i +
B^{-1}K)B^{-1} {\stackrel{({\rm s})}\gamma}$.  Therefore ${\stackrel
  {({\rm s})} \gamma}$\index{Self-dual 2-forms} or, equivalently, 
the {\em triplet of 2-forms}
\begin{eqnarray}
S^{(a)}&:=&-(B^{-1})^{ab}\,\,{\stackrel{({\rm
s})}{\gamma}}{}_b\nonumber\\ &=& {\frac i 2}\,(dx^0\wedge dx^a - (\det
B)^{-1}\,k_b\, dx^b\wedge dx^a \nonumber\\  &&\qquad
+\,i\,(B^{-1})^{ab}\,\epsilon_{bcd}\,dx^c\wedge dx^d).\label{Sa}
\end{eqnarray}  
subsume the properties of this invariant subspace. Each of the 2-forms
carry 3 independent components, i.e., they add up to 9 components.
 
The information of the constitutive matrix $\stackrel{\rm o}
\chi{}^{IJ}$ is now encoded into the triplet of 2-forms $S^{(a)}$.
One can verify that the latter satisfies the completeness 
relation\index{Completeness relation}
\begin{equation}\label{ccr} 
 S^{(a)}\wedge S^{(b)} = {\frac 1 
3}\,(B^{-1})^{ab}\,(B)_{cd}\,S^{(c)}\wedge S^{(d)} \, . 
\end{equation} 
 
\subsubsection{Extracting the metric}
 
Within the context of $SU(2)$ Yang-Mills theory, Urbantke
\cite{Urban1} was able to derive a 4-dimensional spacetime metric
$g_{ij}$ from a triplet of 2-forms satisfying a completeness condition
of the type (\ref{ccr}). Explicitly, the Urbantke formulas\index{Urbantke 
formulas} read
\begin{eqnarray}
  \sqrt{{\det}\,g}\;g_{ij} & =& -\,{\frac 2 3}\,\sqrt{\det 
    B}\,\epsilon_{abc}\, \epsilon^{klmn}\,S^{(a)}_{ik}S^{(b)}_{lm} 
  S^{(c)}_{nj}\,, \label{uf1}\\ \sqrt{{\det}\,g} & =& -\,{\frac 1 
    6}\,\epsilon^{klmn}\,B_{cd}\, 
  S^{(c)}_{kl}S^{(d)}_{mn}\, \label{uf2}.
\end{eqnarray} 
The $S^{(a)}_{ij}$ are the components of the 2-form triplet $S^{(a)} =
S^{(a)}_{ij} dx^i\wedge dx^i/2$. If we substitute the forms (\ref{Sa}) 
into (\ref{uf1}) and (\ref{uf2}), we can display the metric explicitly 
in terms of the constitutive coefficients:
\begin{equation}\label{metric} 
g_{ij} = {\frac 1 {\sqrt{\det B}}}\left(\begin{array}{c|c} \det B & 
-\,k_a \\ \hline -\,k_b & -\,B_{ab} + (\det B)^{-1}\,k_a\,k_b 
\end{array}\right) \, .
\end{equation}  
Here $k_a:=B_{ab}k^b = B_{ab}\,\epsilon^{bcd}K_{cd}/2$. One can
verify that the metric in (\ref{metric}) has Minkowskian signature.
Since the triplet $S^{(a)}$ is defined up to an arbitrary scalar
factor, we obtain a {\em conformal} class of metrics.

Given a metric, we can now define eventually the notion of local
isotropy. Let $T^{i_1\dots i_p}$ be the contravariant coordinate
components of a tensor field and $T^{\alpha_1\dots\alpha_p}
:=e_{i_1}{}^{\alpha_1} \cdots e_{i_p}{}^{\alpha_p}\, T^{i_1\dots i_p}$
its frame components with respect to an orthonormal frame $e_\alpha =
e^i{}_\alpha\,\partial_i$.  A tensor is said to be locally isotropic
at a given point, if its frame components are invariant under a
Lorentz rotation of the orthonormal frame. Similar considerations
extend to tensor densities.

There are only two geometrical objects which are numerically invariant
under local Lorentz transformations: the Minkowski metric
$o_{\alpha\beta}={\rm diag}(+1,-1,-1,-1)$ and the Levi-Civita tensor
density $\epsilon_{\alpha\beta\gamma\delta}$. Thus
\begin{equation}\label{iso} 
{\cal T}^{ijkl}=\phi(x)\, \sqrt{-g}\left( g^{ik}g^{jl} - g^{jk} g^{il}
\right) + \varphi(x)\, \epsilon^{ijkl}
\end{equation}  
is the most general locally isotropic contravariant fourth rank tensor
density of weight $+1$ with the symmetries ${\cal T}^{ijkl}=-{\cal
T}^{jikl} =-{\cal T}^{ijlk}={\cal T}^{klij}$. Here $\phi$ and
$\varphi$ are scalar and pseudo-scalar fields, respectively.

One can prove that the constitutive tensor in (\ref{linear}) with the
closure property (\ref{cr}) is {\em locally isotropic with respect to
  the metric} (\ref{metric}), see also \cite{Ni77}. Accordingly, for
the constitutive tensor, we finally have
\begin{equation}\label{isotropy}
\stackrel{\rm o}{\chi}{}^{\!ijkl}=2\sqrt{-g}\,g^{k[i}g^{j]l}\,.
\end{equation}Thus, the isotropic law reads
\begin{equation}\label{isotropy1}
H_{ij}= \frac{f}{2}\,
\epsilon_{ijmn}\,\sqrt{-g}\,g^{km}g^{ln}\,F_{kl}+ \alpha\,F_{ij}
\end{equation} 
or, if written with the help of the Hodge star operator 
belonging to the metric (\ref{metric}),
\begin{equation}\label{isotropic}
H=\left(f\,^\star+\alpha\right) F\,.
\end{equation}

\subsection{Centrosymmetric}\index{Constitutive law!centrosymmetric}

If we want the constitutive tensor to be reflection symmetric at each
point of spacetime, i.e., if we require centrosymmetry, then we have
to kill the Abelian axion and arrive, provided $F$ is chosen in
accordance with the SI-conventions, at the usual law for
Maxwell-Lorentz vacuum electrodynamics\footnote{Remember that in Ricci
calculus the excitation is defined according to ${\cal H}^{ij}|_{
Ric}=\epsilon^{ijkl}\,H_{kl}/2\,.$ Then ${\cal H}^{ij}|_{
Ric}=-\,f\sqrt{-g}\,F^{ij}$, see (\ref{densities})$_1$.}
\begin{equation}\label{isotropy2}
H=f\,^\star F=\sqrt{\frac{\varepsilon_0}{\mu_0}}\,^\star F\,.
\end{equation}

\section[Non-minimal coupling?]{Non-minimal coupling involving curvature, 
  nonmetricity and torsion?}

\subsection{Non-minimal coupling violating charge and/or flux 
conservation}

The Maxwell equation $d H = J$ reflects (and comes from) the electric
current conservation, $dJ =0$, see Sec.\ref{First}. By modifying the
left hand side of $d H = J$, one can arrive at a model violating
charge conservation. Such a modification can typically originate from
a non-minimal coupling of the electromagnetic to the gravitational
field. Given the torsion 2-form $T^\alpha$\index{Torsion}, one can 
consider, for example, the field equation
\begin{equation}
  dH+\alpha\;\left(e_\alpha\rfloor T^\alpha\right)\wedge H +
  \beta\;{}^\star\!\left(\vartheta_\alpha\wedge T^\alpha\right) \wedge
  H=J\,,
\end{equation} 
or, with the nonmetricity 1-form $Q_{\alpha\beta}
:= - Dg_{\alpha\beta}$\index{Nonmetricity} and the Weyl 1-form 
$Q:=Q_\alpha{}^\alpha/4$,
\begin{equation}
  dH + \gamma\,Q\wedge
  H+\delta\,^\star(\vartheta^\alpha\wedge\vartheta^\beta\wedge
  Q_{\alpha\beta})\wedge H=J\,.
\end{equation} 
Similar non-minimal terms could emerge in $dF=0$. However, curvature
dependent terms cannot be accommodated at the level of the Maxwell
equations, since the contraction of the indices produces always a form
of even rank whereas the Maxwell equations are represented by 3-forms,
i.e., by forms of odd rank. In any case, violating charge or flux
conservation is not possible without giving up most of the
experimentally established structure of the theory of
electromagnetism. Therefore we will not follow this path.

Incidentally, there are some papers in the literature in which the
conventional vacuum constitutive law $H\sim{}^\star F$ is uphold, but
the Maxwell equations are coupled to torsion in an inconsistent way. A
closer inspection of the papers
\cite{Prasanna75,Ruggiero83,Larry86} shows that the proposed
``non-minimal" coupling of torsion to the electromagnetic field is
void of physical contents. In fact, torsion drops out if the algebra
is done correctly.

Another procedure comes to mind if we talk about the violation of the
conservation laws. Hojman et al.\ \cite{Hojman} introduced a new
scalar field $\varphi(x)$, the {\em tlaplon}, see also Mukku \& Sayed
\cite{Mukku}. The gradient $d\varphi$ of the tlaplon
 was put proportional to the trace part $^{(2)}T^\alpha:=
\vartheta^\alpha\wedge T$ of the torsion
$T^\alpha$, see \cite{PRs}; here $T:=e_\beta\rfloor T^\beta$. In fact,
we have $T=\frac{3}{2}\,d\,\varphi$.

Superficially, the axial scalar $\alpha(x)$ (Abelian axion) and the
scalar tlaplon\index{Tlaplon} $\varphi(x)$ may look similar. However, the axion
already emerges from spacetime viewed as a differential manifold as
soon as a linear constitutive law is assumed for electromagnetism,
whereas the tlaplon can only be introduced if the differential
manifold in equipped with a linear connection. In other words, the
axion is a pre-metric and a pre-connection animal, the tlaplon, in
contrast, needs to be `housed' in a linear connection.

Moreover, as we saw, the axion respects the conservation laws (and
pleases us thereby), whereas the tlaplon defies these rules and
appears as an anti-electromagnetic creature. The electromagnetic field
is not defined in the conventional way, namely by $F = dA$. Instead,
Hojman et al.\ define it via a ``covariant" derivative according to
\begin{equation}
\widehat{F} = F - T^\alpha\,A_\alpha = F + {\frac 1 3}\,
T\wedge A = F + {\frac 1 2}\,d\varphi\wedge A.
\end{equation}
Thus the ``electromagnetic'' Lagrangian becomes
\begin{equation}\label{tlaplonLagrangian}
\widehat{F}\wedge{}^\star\widehat{F} = F\wedge{}^\star F +
d\varphi\wedge A\wedge{}^\star F +
{\frac 1 4}\,T\wedge A\wedge{}^\star(T\wedge A)\,.
\end{equation}
The last term does not contribute to the excitation $H = - \partial
L/\partial F$, but the second term produces a contribution
\begin{equation}
\sim {}^\star (d\varphi\wedge A).\label{tor-mom}
\end{equation}

We can compare now the two contributions from the axion (\ref{ax-mom})
and the tlaplon (\ref{tor-mom}). They are reminiscent of each other
since one is equal to the Hodge dual of the other. Therefore, in the
tlaplon case, besides a connection, we need additionally a metric. But
the most decisive difference is, as can be read off from
(\ref{tlaplonLagrangian}), that the Maxwell equations get amended and
axiom 1 and axiom 3 are no longer valid.

\subsection{``Admissible'' non-minimal coupling}

The message is then that a change of the Maxwell equations
$dH=J\,,\;dF=0$ is to be avoided, unless one allows for a violation of
electric charge or magnetic flux conservation. By introducing the
metric $g^{ij}$ into the constitutive law, one gets a smooth and
natural transition from special relativity to general relativity and
to gauge theories of gravity. In the constitutive law for vacuum one
could imagine, along with the contributions depending only on the
metric, couplings like \cite{Hubert1}\index{Non-minimal coupling}
\begin{eqnarray}
\chi^{ijkl} &=& A_1\,R^{ijkl} + A_2\,R^{\ast\,ijkl} + A_3\,^\star\!
R^{\ast\,ijkl} \nonumber\\
&& +\,A_4\,({ Ric}^{i[k}g^{l]j} - { Ric}^{j[k}g^{l]i}) 
+ A_5\,R\,g^{i[k}g^{l]j} + A_6\,R\,\epsilon^{ijkl}\label{RFcoupling}
\end{eqnarray} 
without violating the conservation laws. Here $R:=e^\alpha\rfloor
e_\beta\rfloor R_\alpha{}^\beta$ is the curvature scalar, and we
denote the {\it right} or {\it Lie} dual of an $so(1,3)$-valued form
$\psi_{\alpha\beta}$ by $\psi^\ast_{\alpha\beta}:={\frac 1 2}\,
\epsilon_{\alpha\beta\mu\nu}\,\psi^{\mu\nu}$.

Choose, for example, $\chi^{ijkl} = 2\sqrt{-g}f_0(1 +
\beta^2\,R)\,g^{i[k}g^{l]j}$, then the inhomogeneous Maxwell equation
would read
\begin{equation}\label{RFcoupling1}
  d{}\,\,^\star\! F+\beta^2d\left(R{}\,\,^\star\! F\right)=\frac{1}{f_0}\,J\,,
\end{equation}
i.e., a coupling of curvature and electromagnetic field strength would
be possible.  However, one had to introduce a new natural constant
with the dimension of $[\beta]=1/{\rm length}$. On the level of the
Lagrangian, this coupling would be non-minimal,
\begin{equation}\label{RFcoupling2}
V_{\rm non-m}
=-\,\frac{f_0}{2}\,\left(1+\beta^2R\right){}^\star\!F\wedge F\,,
\end{equation}
but such an ansatz would not spoil the fundamental principles of
electrodynamics; it would seem to be the most natural way of achieving
an $R\,F$-coupling. Goenner \cite{Hubert1}, see also the literature
given there, derived this non-minimal Lagrangian from some fundamental
principles, like the existence of a decent Newtonian limit. However,
in his view, such a model violates charge conservation. We disagree
with him on this point.

We want to stress that one cannot achieve a similar non-minimal
coupling to the torsion $T^\alpha$ of spacetime. First of all, the
Maxwell equations are independent of torsion. By means of the
constitutive law one maps the 2-form $F$ to the 2-form $H$. The
curvature $R_\alpha{}^\beta$ is a 2-form of type $(1,1)$, i.e., it
carries two $GL(4,R)$ indices, whereas the torsion 2-form carries only
one index. Therefore, by contraction, we cannot get a scalar out of
the torsion. A coupling like $\left(e_\alpha\rfloor T^\alpha
\right)\wedge {}^\star\!F$ is not possible, since this is a 3-form.
However, higher powers in $T^\alpha$\index{Torsion} would be possible 
such as
\begin{equation}\label{nmtorsion}
H=f_0\left[1 + \gamma^2{}\,^\star\!\left(T^\alpha\wedge
T_\alpha\right)\right] \wedge {}^\star\!F\,
\end{equation}
or 
\begin{equation}\label{nmtorsion1}
H=f_0\left[1  + \delta^4{}\,{}^\star\!\left(
(e_\alpha\rfloor T^\alpha)\wedge (e_\beta\rfloor
T^\beta)\wedge (e_\gamma\rfloor T^\gamma)\wedge
(e_\delta\rfloor T^\delta)\right)\right]\wedge{}^\star\!F\,.
\end{equation}
Here $[\gamma]=[\delta]=1/length$. Accordingly, it is not too
difficult to introduce a coupling of torsion to electromagnetism.
However, the price one has to pay is the introduction of new natural
constants $\gamma$ and $\delta$. In other words, even if possible, we
don't take such models too seriously.

Also non-minimally coupled nonmetricity\index{Nonmetricity} could be 
installed by additional quadratic pieces such as
\begin{equation}\label{nonminimalQ}
  H=f_0\left[1+\xi^2(e_\alpha\rfloor Q^{\alpha\gamma})(e_\beta\rfloor
    Q^{\beta}{}_\gamma) \right]{}^\star \!F \,.
\end{equation} 
Therefore, there are quite a number of different ``admissible''
options available as soon as we allow non-minimal couplings to arise.

\section{Outlook}

Using astronomical observations on the propagation of light, the upper
bounds for non-minimal coupling effects should be determined in a
systematic way, as is done, for example, by Haugan and L\"ammerzahl
\cite{HLaem}. For such a purpose, we will develop \cite{geomopt} the
geometrical optics limit of the Maxwell equations $dH=J,\,\, dF=0$ and
will use particular constitutive laws, as, e.g., the linear law.
Possible couplings to curvature, torsion, and nonmetricity should come
under sharper focus in this way. Non-linear effect \`a la Ay\'on-Beato
\& Garc{\'\i}a should be investigated in the context of, say, the
metric-affine gauge theory of gravity.  \bigskip

{\bf Acknowledgments:} The authors are indebted to Steve Adler
(Princeton), Tetsuo Fukui (Nishinomiya), Hubert Goenner (G\"ottingen), 
Yakov Itin (Jeru\-salem), Claus L\"am\-mer\-zahl (Kons\-tanz), Bah\-ram 
Mash\-hoon (Co\-lumbia, MO), and Guillermo Rubilar (Cologne) for helpful 
remarks and/or interesting discussions. FWH is grateful to the Institute 
for Advanced Study for hospitality and to the Volks\-wagen-Stift\-ung, 
Hannover for support.

\begin{footnotesize}

\end{footnotesize}

\end{document}